# A Dual Heterogeneous Island Genetic Algorithm for Solving Large Size Flexible Flow Shop Scheduling Problems on Hybrid multi-core CPU and GPU Platforms

Jia Luo*(jluo@laas.fr), Didier El Baz

LAAS-CNRS, Université de Toulouse, CNRS, Toulouse, France

## Abstract

The flexible flow shop scheduling problem is an NP-hard problem and it requires significant resolution time to find optimal or even adequate solutions when dealing with large size instances. Thus, this paper proposes a dual island genetic algorithm consisting of a parallel cellular model and a parallel pseudo model. This is a two-level parallelization highly consistent with the underlying architecture and is well suited for parallelizing inside or between GPUs and a multi-core CPU. At the higher level, the efficiency of island GAs is improved by exploring new regions within the search space utilizing different methods. In the meantime, the cellular model keeps the population diversity by decentralization and the pseudo model enhances the search ability by the complementary parent strategy at the lower level. To encourage the information sharing between islands, a penetration inspired migration policy is designed which sets the topology, the rate, the interval and the strategy adaptively. Finally, the proposed method is tested on some large size flexible flow shop scheduling instances in comparison with other parallel algorithms. The computational results show that it cannot only obtain competitive results but also reduces execution time.

## Key words

Heterogeneous island genetic algorithm

Heterogeneous architectures

Multi-core processing

GPU computing

Flexible flow shop scheduling

---

\* Corresponding author

# 1. Introduction

The Flexible Flow Shop scheduling problem (FFS) focuses on improving machine utilization and reducing make-span. Some works on solving small size FFS are concerned on exact methods [1][2] to find the optimal solutions. However, conventional optimization techniques always fail in industry application as the problem sizes in the real world are much bigger and the computational cost is increased. Therefore, there is a growing interest in developing heuristic methods to solve large complex FFS problems [3][4]. Although these approaches cannot guarantee finding optimal solutions, there is a sizable probability that an adequate solution is found in a reasonable time.

The Genetic Algorithm (GA) is one of the most widely known heuristic methods and is one of the best approaches in solving FFS problems. But when GAs are applied to large or complex problems, there is a conflict between searching better solutions and execution time. In contrast to classical GAs, the island GA [5] divides the population into a few relatively large subpopulations. Each of them works as an island and is free to converge towards its own sub-optimum. At some points, a migration operator is utilized to exchange individuals among islands. This imitates the nature in a better way which increases the search diversification [6]. Furthermore, it is one of the most famous models to exploit parallelism in GAs. Nevertheless, due to the same genetic operator configurations in each island, island GAs are apt to yield premature convergence. Meanwhile, this design has to be carried out with respect to the underlying architectures for parallelization implementation.

With the unprecedented evolution of GPUs and multi-core CPUs, almost all modern computers are equipped with both. Some comparisons between their performances for GA applications were discussed [7], but the cooperation between the two in this domain was rarely concerned. These facts have motivated the design of a heterogeneous island GA that keeps better population diversity and is well suited for parallelization on GPUs and a multi-core CPU. In this paper, we seek to address it and its application to a large size FFS problem. Specially, the contributions of our work are summarized as follows:

1. a dual heterogeneous island model is proposed where the 2D variable space of the cellular GA and the complementary parent strategy of the pseudo GA keep the population diversity;

2. a two-level parallelization highly consistent with the underlying architecture is implemented that is well suited for parallelizing inside or between GPUs and a multi-core CPU;

3. a penetration inspired migration policy is designed so that it can share good individuals effectively by setting the topology, the rate, the interval and the strategy adaptively.

The remaining sections of this paper are organized as follows. Section 2 introduces related works. Section 3 describes the research problem. Section 4 presents the design of the dual heterogeneous island GA on hybrid multi-core CPU and GPU platforms. Section 5 presents the numerical experiments and result analysis. Finally, section 6 states the conclusions.

## 2. Related Works

When the population size is N and there are n islands, only N/n individuals work with GA operators in one island. Moreover, the selection and the elitist strategy in GAs decrease the subpopulation diversity in one island after several generations. Although the migration at some points can help create new individuals, the influence is restricted because GA operators in each island function in the same way. What is worse, an inappropriate implementation of migration mechanism may cause genetic drift and leads to converge toward a local optimum. One approach for dealing with this problem is the heterogeneous island GA which makes distinction among subpopulations by different configurations. Herrera et al. [8] presented the gradual distributed real-coded GA that applied different crossover operators to different subpopulations. Alba et al. [9] encompassed the actual parallelization of the gradual distributed real-coded GA on a cluster of 8 homogeneous PCs. In [10], Miki et al. designed a parallel GA using nCUBE-2E where different islands had different parameter settings. Although these heterogeneous algorithms have improved the solutions' quality, the implementation is usually executed on a homogeneous architecture or even on a mono processor. In these cases, different islands can work in parallel but GA operations inside one island are executed in a sequential way.

In addition to propose heterogeneous island GAs, some works were carried out to evaluate the performance of heterogeneous computing architectures for island GAs. In [11], a homogenous island GA was run at the same time on different types of machines which obtained super-linear speedup. García-Sánchez et al. [12] studied benefits from setting the subpopulation sizes according to each heterogeneous node's computational power. García-Valdez et al. [13] tested the randomized parameter setting strategy for heterogeneous workers in pool-based GAs. Despite promising results from leveraging computational capabilities of a heterogeneous cluster, these methods must face some common challenges such as lost connections, low bandwidth, abandoned work, security and privacy. Moreover, the above-mentioned designs generally are hard to profit the computation capability from GPUs or heterogeneous environment mixed with multi-core processors and many-core processors.

Since the cooperation between GPUs and a multi-core CPU is stable and secure, some efforts have considered to utilize both and enjoy their compute capabilities maximally. Dabah et al. [14] proposed 5 accelerated branch and bound algorithms for solving the blocking job shop scheduling problem where two of them presented a hybridization between the multi-core CPU approach and the GPUs-based parallelization approach. Benner et al [15] discussed a hybrid Lyapunov solver where the intensive parts of the computation were accelerated using GPUs while executing the remaining operations on a multi-core CPU. In [16], Bilel et al. introduced a CPU-GPU co-simulation framework where synchronization and experiment design were performed on CPU and node's processes were executed in parallel on GPUs. These studies have confirmed the interest to design a scheme that exploits GPUs and a multi-core CPU in efficient ways. However, simultaneous parallelization on two sides and its implementation for island GAs are not yet concerned.

Several researches have tried island GAs to solve shop scheduling problems either for improving the solutions' quality [17][18] or for decreasing the execution time [19][20]. But none of them have so far, and to the best of our knowledge, considered heterogeneous island GAs parallelized on GPUs and a multi-core CPU. All the above-mentioned efforts provide us a starting point for designing a dual heterogeneous island GA that keeps a better population diversity and that is well suited for parallelization on hybrid multi-core CPU and GPU platforms.

## 3. Problem Definition

The FFS is a multistage production process as illustrated in Figure 1. An instance of the FFS problem considers a set of J jobs ($1 \leq j \leq J$). Each of them consists of a set of S stages ($2 \leq s \leq S$). At every stage, there is a set of $M_s$ machines ($1 \leq m \leq M_s$) and at least one stage has more than one machine. All jobs need to go through all stages in the same order and only one machine is selected for processing on each stage. There is no precedence between operations of different jobs, but there is precedence among operations due to the jobs' processing cycles. Preemptive operations are not allowed. A feasible solution is described by jobs' sequence on target machines $M_{js}$. The processing time of job j at stage s on machine m is abbreviated as $P_{jsm}$. Usually, it is known with the release time $R_j$ and the due time $D_j$. The objective function to minimize the total tardiness and the makespan is represented by $WT * \sum T_j + C_{max}$ using the classification scheme of Bruzzone et al. [21], where WT indicates the priority of the first objective. As a minimization problem, the fitness function of an individual is transferred from the objective function by $\max(E_{max} - (WT * \sum T_j + C_{max}, 0)$, where $E_{max}$ is the estimated maximum value of the objective function. The FFS problem is NP-hard in essence and is thus difficult to solve [22]. When dealing with large size instances, it requires huge resolution time to find optimal or even adequate solutions.

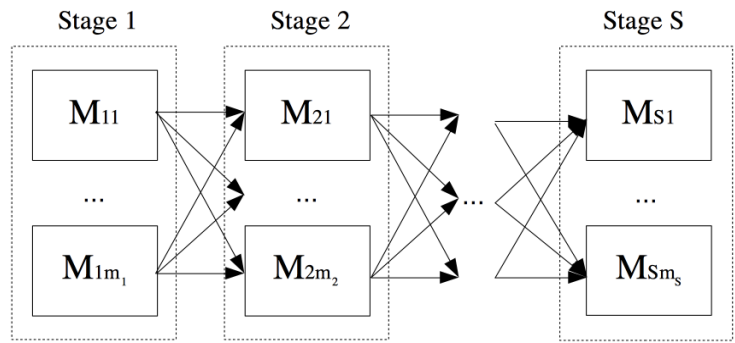

Fig.1. A flexible flow shop layout

# 4. Dual Heterogeneous Island Genetic Algorithm on Hybrid multi-core CPU and GPU Platforms

4.1 Dual Heterogeneous Island Strategy

The general framework of the proposed dual heterogeneous island strategy is shown in Fig. 2. There is the same number of individuals on each island where island A works with the cellular GA [23] and island B works with the pseudo GA [24]. As two islands are exploring new regions within the search space utilizing different methods, it helps enlarge the scope of the search process and increase the chances of avoiding premature convergence. Moreover, individuals from heterogeneous islands have obtained different characters during the independent evolution procedure. In this case, the benefit of migration is increased. At the software level, three sublevels are considered according to the source of the heterogeneity:

- Genotype level: As a feasible solution is described by jobs' sequence on target machines, the chromosome is displayed by a string of length $J{\times}S$ and is indexed from 0 to $J{\times}S-1$. The i-th gene states the index of the target machine for job $\lfloor i/S \rfloor + 1$ at stage $\{i/S\}+1$ and each gene has two layers. The upper layer is designed for the cellular GA where the i-th gene is presented by an integer number. At the lower layer, the i-th gene is expressed by binary numbers to work with the complementary parent strategy of the pseudo GA.

- Operator level: The cellular GA starts with random initialization and maps individuals on a 2D grid. An individual is limited to compete and mate with its neighbors, while the neighborhoods overlapping makes good solutions disseminate through the entire population. This design allows a better exploration of the search space with respect to decentralization. The pseudo GA initializes every pair of parents by the dynamic complementary strategy [24]. The evolution is executed between the offspring from the same parents, during which the parents are completely replaced by their own children. In this case, search ability is enhanced since higher population diversity is got without gene lost.

- Parameter level: The execution of the crossover operator and the mutation operator are determined by the crossover rate and the mutation rate. Their values for the cellular GA and the pseudo GA on different islands are set differently.

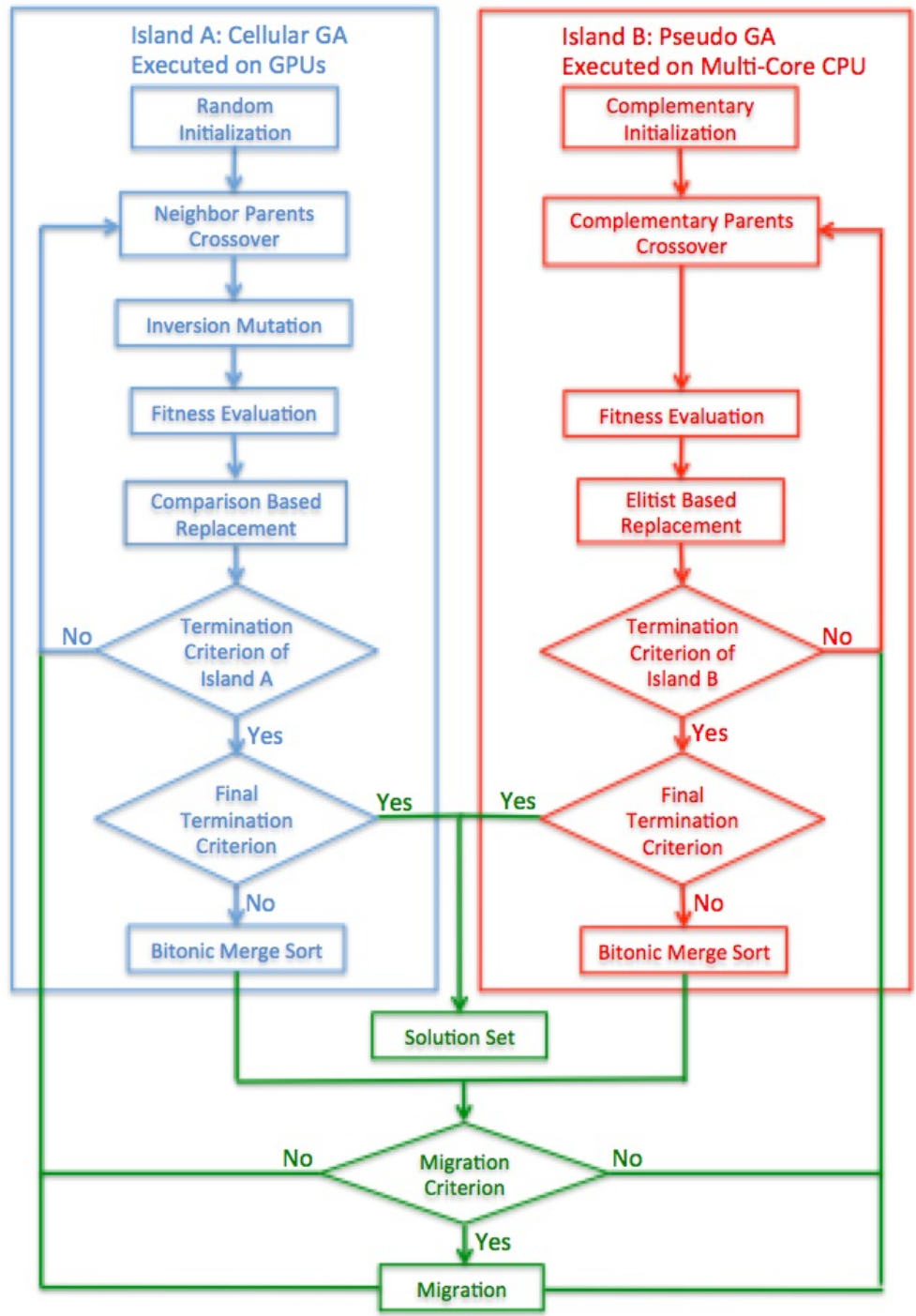

Fig.2 The general framework of the dual heterogeneous island GA

4.2 Parallelization on Hybrid multi-core CPU and GPU Platforms

As far as the hardware level is concerned, there are two obvious advantages to parallelize the dual heterogeneous island GA on hybrid multi-core CPU and GPU platforms:

- Widespread HPC resources: Nowadays, almost all modern computers are equipped with GPUs and a multi-core CPU. The cooperation between them is through their inner connections which is stable and secure. With the development of CUDA [25], it is convenient to use enabled GPUs for general purpose processing. On the other hand, concurrency platforms allowing the coordination of multicore resources facilitate programming on multi-core CPUs. Moreover, in addition to the parallelization on GPUs or on a multi-core CPU at the lower level, the GPUs and the multi-core CPU can work concurrently at the higher level to maximally use computing resources.

- High consistency with the proposed GA: The cellular GA maps individuals on a 2D grid and the CUDA threads are grouped into 2D blocks that are organized in a 2D grid, using the local memory, the shared memory and the global memory respectively [26]. Thus, the cellular GA can be entirely parallelized on GPUs. On the other hand, only the crossover, the fitness evaluation and the replacement are kept in the pseudo GA. The crossover is performed between fixed complementary parents. The fitness evaluations of individuals are independent. Since no global information is required, all for loops in the above two steps can be easily handled on a multi-core CPU in parallel.

As the texture caches of CUDA are designed to gain an increase in performance for accelerating access patterns with spatial locality [27], we design the neighborhood area of the cellular GA as shown in Fig. 3. Individuals' information and GA operators are placed and executed through the global memory while the neighbors' information are stored in the texture memory. Each CUDA thread handles one cellule of the cellular GA. Firstly, it recombines two individuals selected from the nearby area to generate a new one. Afterwards, this new individual undertakes the mutation and replaces the original individual if its solution is better. Then, all individuals are sorted according to their fitness values using the Bitonic-Merge sort [28], if the cellular GA meets the island termination criterion but not the final termination criterion.

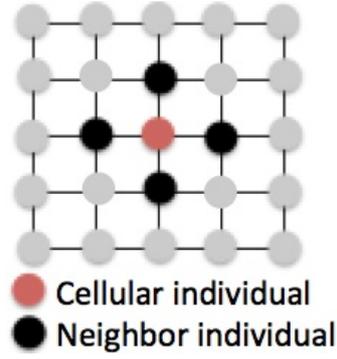

Fig.3 The neighborhood area of the cellular GA

When the GPUs are occupied by executing the cellular GA, the pseudo GA is run on a multi-core CPU by OpenMP [29] which is an API supporting multi-platform shared memory multiprocessing programming. In this case, the GA operators on two heterogeneous islands are working in parallel on the host (a multi-core CPU) and the device (GPUs) simultaneously. At the end, the Bitonic-Merge sort [28] is accomplished by the OpenMP-based code in a similar way as the cellular GA on CUDA.

4.3 Migration Policy

The migration between islands is controlled by the topology, the rate, the interval and the strategy. To decrease the number of parameters that need to be set manually, we develop a migration policy inspired by the penetration theory [11] where a migration threshold value $\theta$ is set ($0 \leq \theta \leq 1$). The execution of migration is decided by this value and there is more likely for individuals to migrate when $\theta = 1$. Moreover, the migration rate $\alpha$ and the migration direction indicator $\beta$ are formulated as in equation (1) and equation (2), respectively:

$$\alpha = \begin{cases} 1 - \beta & 1 - \beta < \theta \\ 0 & 1 - \beta \geq \theta \end{cases} \quad (1)$$

$$\beta = \begin{cases} \text{fit}_A/\text{fit}_B & \text{fit}_A < \text{fit}_B \\ \text{fit}_B/\text{fit}_A & \text{fit}_A > \text{fit}_B \end{cases} \quad (2)$$

Here, $\text{fit}_A$ and $\text{fit}_B$ are the best individual's fitness value of subpopulation A on island A and subpopulation B on island B. In a certain generation, we calculate the above functions and carry out three steps as follows:

- If $1 - \beta < \theta$, the migration is executed. Otherwise, do nothing.

- The topology of migration is determined by the ratio of $fit_A$ to $fit_B$. If $fit_A/fit_B > 1$, the migration is from subpopulation A to subpopulation B. If $fit_A/fit_B < 1$, the migration direction is reversed. If $fit_A/fit_B = 1$, no migration is implemented.

- When the migration is carried, $\alpha \times N$ individuals with best fitness values in the emigrant subpopulation are selected to replace $\alpha \times N$ individuals with worst fitness values in the immigrant subpopulation.

The migration policy is executed by the CPU where results of cellular GA on GPUs are sent back to the CPU at this moment. With this design, the topology, the rate, the interval and the strategy no longer need to be considered manually. New merged individuals with good genes can be transited quickly and the execution time is saved by preventing ineffective information sharing.

## 5. Numerical Experiments

To analyze the performance of the proposed algorithm, we compare its solutions' quality and execution time with the parallel cellular GA and the parallel pseudo GA. The population size is kept as 512 for all tested GAs while the subpopulation size for each island of heterogeneous GA is 256. The crossover rate and the mutation rate of cellular GA are set as 1.00 and 0.05 respectively [23], while the crossover rate of pseudo GA is equal to 0.75 [24]. The cellular GA from the dual heterogeneous GA keeps the same crossover rate and mutation rate as the cellular GA. Similarly, the pseudo GA from the dual heterogeneous GA keeps the same crossover rate as the pseudo GA. Moreover, to better check the influence of migration, the migration threshold is fixed as 1.00. As a large size FFS is concerned in this paper, all analyzed instances are characterized by 300 jobs with 4 stages and there are 2 available machines at each stage. Other experimental relative data are defined in Table 1.

Table 1. The experimental relative data of the large FFS problem

| | |
|---|---|
| WT | 100 |
| $P_{jsm}$ | $U[1, 5]$ |
| $R_j$ | $U[0, \overline{P}]$, where $\overline{P}=\sum_j \sum_s (\sum_m P_{jsm}/M_s)$ |
| $D_j$ | $R_j + \overline{P}_j(1 + \sigma)$, where $\sigma=U[0,2]$ and $\overline{P}_j = \sum_s (\sum_m P_{jsm}/M_s)$ |

The experimental platform is based on the Intel Xeon E5640 CPU with 2.67GHz clock speed and four cores. The GPU code implementation is carried out using CUDA 8.0 on NVIDIA Tesla K40, with 2880 cores at 0.745 GHz and 12 GB GDDR5 global memory. All programs are written in C, except for the GPU kernels in CUDA C. The following table and figures display results of 2000 generations and they are average values of 50 runs.

5.1 Test on Migration Policy Execution Gap

Even the topology, the rate, the interval and the strategy are set adaptively when the migration policy is carried in a certain generation. We still need to test when to execute it since the migration policy needs call back results on GPUs and too frequent data exchange between the device and the host may weaken the performance of the proposed method. As it is displayed in Fig.4, the migration policy execution gap is increased from 10 generations to 800 generations and the island GA has a risk to fall in a local optimum if this value is either too small or too big. As a result, it finds that an inappropriate migration can also lead onto premature convergence, besides homogeneous genetic operator configurations and limited subpopulation sizes. Following the polynomial fitting values, the best performance for the tested instance is obtained when the migration policy execution gap is around 500 generations and we keep this setting for the remaining tests in this paper.

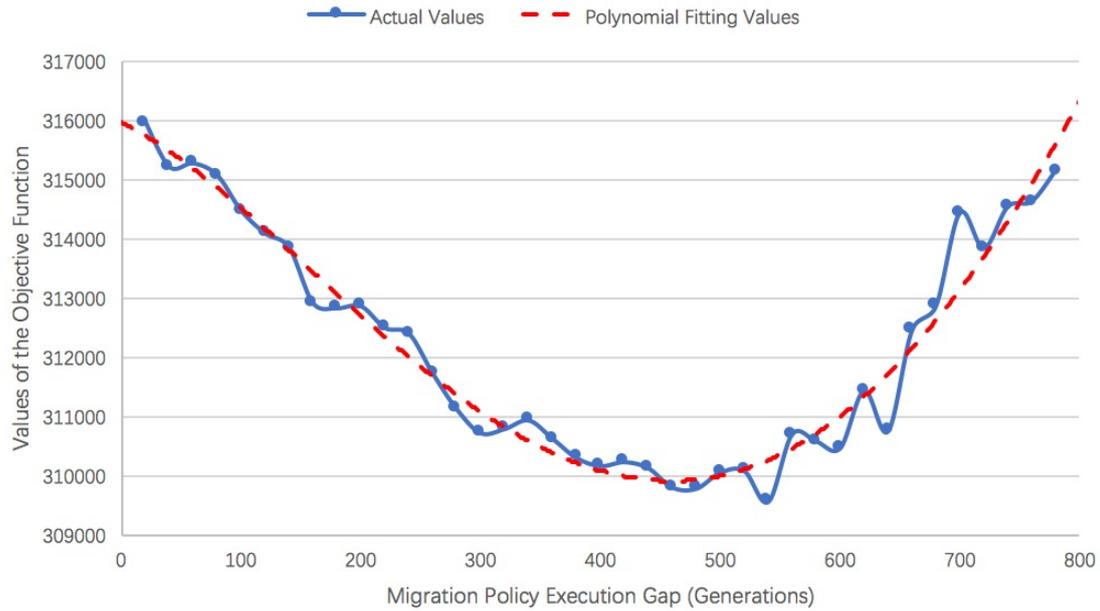

Fig.4 The influence of the migration policy execution gap for the heterogeneous GA

5.2 Comparison Test on Solutions' Quality

The solutions' quality of different GAs are shown in Table 2. Although the specific designs of cellular GA and the pseudo GA can help increase population diversity, the proposed method combines the merits from both and optimizes the performance by independent evolution and penetration migration. Thus, the heterogeneous GA overcomes them with better solutions and less variance. This effect is also confirmed by the convergence trend among three GAs in Fig. 5. Moreover, there are elbows in the convergence curve of the designed approach and they always appear around the generations where the migration policy is executed. This phenomenon witnesses the process of how the premature convergence is avoided thanks to two heterogeneous islands connected by the penetration migration.

Table 2. The solutions' quality comparison among different GAs

| Different GAs | Best | Average | Variance |
| --- | --- | --- | --- |
| Heterogeneous GA | 306500.03 | 309885.90 | 2003059.14 |
| Cellular GA | 314467.50 | 320648.18 | 6792896.04 |
| Pseudo GA | 314636.59 | 317683.23 | 2963668.96 |

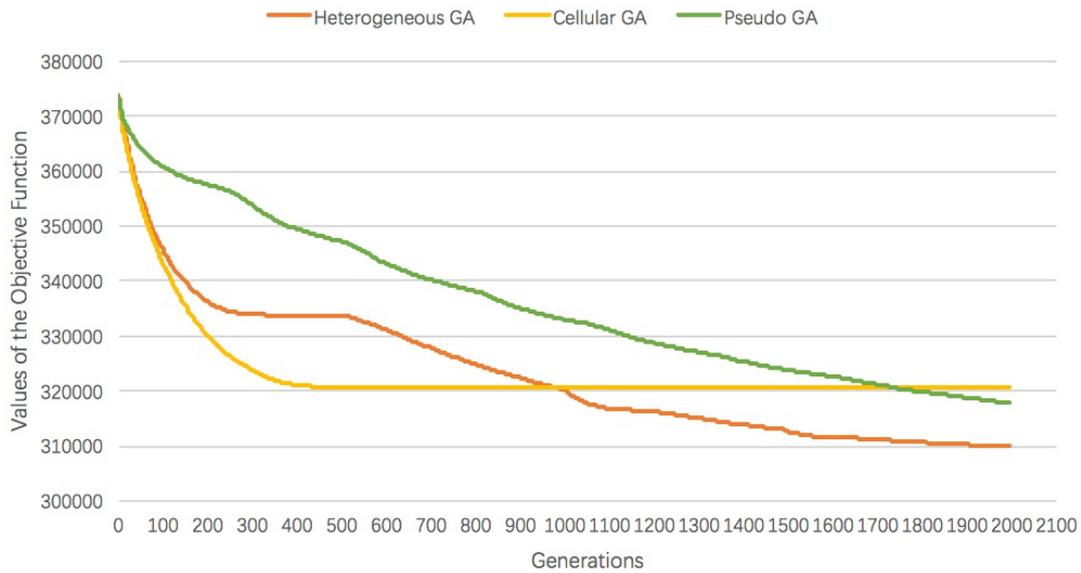

Fig.5 The convergence trend among different GAs

5.3 Comparison Test on Execution Time

To check the execution time among these parallel GAs, we consider different population sizes from 512 to 4096. The cellular GA is fully carried on GPUs. The pseudo GA is generated on a four-core CPU with or without SIMD vectorization. The two islands of heterogeneous GA are generated on GPUs and a CPU simultaneously. Similarly, the pseudo GA from the dual heterogeneous GA is parallelized on the 4 core CPU with or without SIMD vectorization. The SIMD vectorization is executed via SSE2 [31], as far as this experiment platform is available. Concerning results in Fig. 6, the heterogeneous GA on the hybrid platform takes less execution time than the pseudo GA on a 4 core CPU as the heterogeneous design can be well parallelized on both sides simultaneously. However, it loses to the cellular GA because the amount of individuals executed on GPUs and the threads occupancy are twice as much as the heterogeneous GA on the hybrid platform. Fortunately, the performance of the heterogeneous structure gets improved significantly when the computation capability on the 4 core CPU is enhanced by the SIMD vectorization. It points out the importance of computation capability balance between the host and the device when the proposed approach is implemented where the weak side may become as a bottleneck and reduces the overall effectiveness. Finally, because the pseudo GA only deals only with binary integers whose storage size is small, the contribution of SIMD vectorization is impressive and the pseudo GA on a 4 core CPU with vectorization overcomes the others.

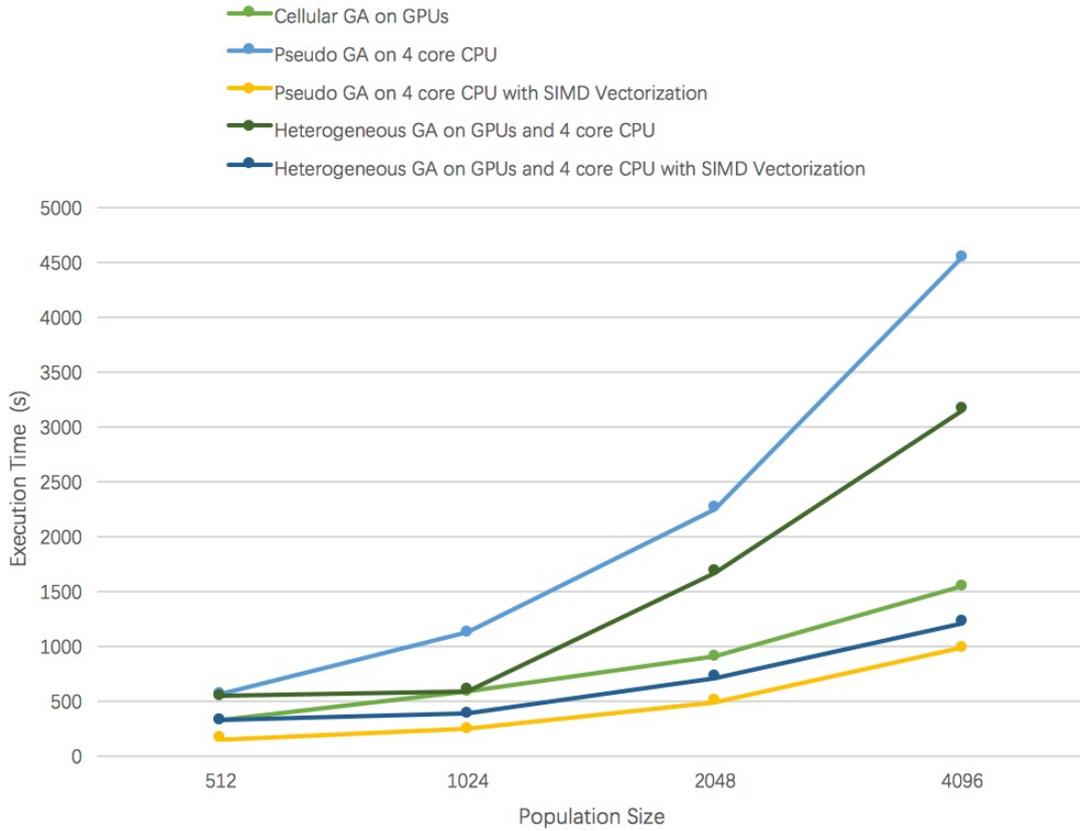

Fig.6 The execution time comparison among different parallel GAs

## 6. Conclusions

A dual heterogeneous island GA was proposed in this paper. It was composed of a cellular GA on GPUs and a pseudo GA on a multi-core CPU where the 2D variable space of the cellular GA and the complementary parent strategy of the pseudo GA kept the population diversity. This structure was highly consistent with the underlying architecture which can be parallelized inside or between GPUs and a multi-core CPU. Since the two islands evolved independently in different ways, a penetration inspired migration was designed to share information between them and to decrease the risk of premature convergence. For solving some large instances of the FFS problem, it firstly found out the importance of an appropriate migration implementation. Otherwise, the migration could cause genetic drift and lead to a convergence towards a local optimum. The second test showed the proposed method obtained better solutions with less variance because of the merits from two different islands and confirmed the efficiency of the penetration migration. Finally, the effectiveness of the dual heterogeneous island GA was displayed by comparison tests with other parallel methods and pointed that the balance of computation capability between the host and the device had a great influence on its overall performance.

# Acknowledgement

This work was supported by a scholarship from the China Scholarship Council (CSC).